\documentclass[epj]{svjour}
\usepackage{graphics}
\begin{document}
\title{Masses of tetraquark states in the hidden charm sector above $D - D^*$ threshold}
\author{Tanvi Bhavsar\inst{1}, Manan Shah\inst{2},  Smruti Patel \inst{3} \and  P. C. Vinodkumar\inst{1}
\thanks{\emph{Present address:} tanvibhavsar1992@yahoo.com}%
}                     
\offprints{}          
\institute{Department of Physics, Sardar Patel University,Vallabh Vidyanagar, INDIA. \and P. D. Patel Institute of Applied Sciences, CHARUSAT, Changa, 388421, India. \and Government Science College,songadh, 394670, India.}
\date{Received: date / Revised version: date}
%
\abstract{
In the diquark-diantiquark composition, we study the masses of hidden charm tetraquark systems ($cq\bar{c}\bar{q}$, $cs\bar{c}\bar{s}$ and $cs\bar{c}\bar{q}$) using a linear confinement potential. In this study, we have factorized the four body system into three subsequent two body systems. To remove degeneracy in the S and P wave masses of mesons and tetraquark states, the spin-spin, spin orbit and tensor components of the confined one gluon exchange interactions are employed. In this attempt, we have been able to assign the $\psi(4230)$  as pure $cq\bar{c}\bar{q}$ tetraquark state. $\psi(4360)$ and $\psi(4390)$ as pure $cs\bar{c}\bar{q}$  tetraquark states.  According to our analysis $\psi(4260)$ is an admixture of  $^1P_1$ and  $^5P_1$ $cq\bar{c}\bar{q}$ tetraquark state. Additionally, we have been able to predict the radiative decay width  $\Gamma_{(\psi \rightarrow J/\psi \gamma)}$, leptonic decay width $\Gamma_{e^+e^-}$ and hadronic decays of $1^{--}$ tetraquark states.
} 
\maketitle
\section{Introduction}
\label{intro}
The study of hadron physics was simple before 2003 because we had a very successful description of the structure of hadrons. However, after the discovery of X(3872) by Belle collaboration in 2003 \cite{belle} many unexpected exotic hadrons such as the charmonium -like and bottomonium-like X Y Z states, hidden-charm pentaquarks,  hybrid mesons etc., have been reported experimentally. These X Y Z states are termed as exotics as they do not fit into the conventional mesons (quark-antiquark bound system) or baryons (three quark system) of the quark model. In the past, the existence of multiquark states has been predicted theoretically by many. In 1977, R. L. Jaffe proposed a radical solution for
the $Q^2$$\bar{Q}^2$ problem and they have examined the S-wave $Q^2$$\bar{Q}^2$ sector using the MIT bag model \cite{rlj}. I. M. Barbour and D. K. Ponting used variational method to solve the $Schr\ddot{o}dinger$ equation for the nonrelativistic $qq$$\bar{q}$$\bar{q}$ system and they have predicted the mass spectrum for the non-strange baryonium state in 1979 \cite{imb}. L. Heller and J. A. Tjon have used the potential energy coming from the MIT bag model to describe a four quark system \cite{lh}. In their study, they have taken the lower eigenvalue of the 2 $\times$ 2 potential matrix and have variationally solved the four-body Schrodinger equation and  predicted the bound states of dimesons [$c^2$$\bar{c}^2$ and $b^2$$\bar{b}^2$] \cite{lh}. B. Silvestre-Brac and C. Semay have studied all possible diquarkonia ($q^2\bar{q}^2$) for different flavours (u, d, s, c, b) for total spin S, for an orbital angular momentum L $<$ 4 using non-relativistic quark model \cite{bsb,bhaduri}.\\

The recent experimental discovery of many \textit{ X Y Z } and $\textit{P}_c$  along with other exotic charged states which are observed by experimental collaborations such as LHCb, Belle, CDF, D0, BESIII, BABAR and CLEOc. Charged states such as $Z_c(3900)$ \cite{p,q,r,s}, $Z_1(4050)$ \cite{z}, $Z_c(4200)$ \cite{t}, $Z_2(4250)$ \cite{z} have renewed interest in the exotic hadronic sector. Very recently, Jing Wu et al. have Systematically studied the exotic hadronic states using chromomagnetic interaction (CMI) model \cite{jw}. They have predicted that X(4140) is the lowest $cs\bar{c}\bar{s}$ state having $J^{PC}$ = $1^{++}$ and have estimated the masses of the other tetraquark states \cite{jw}. Zhi-Gang Wang studied the vector tetraquark states with the QCD sum rules, and they have predicted the status of  Y(4220/4260), Y (4320/4360), Y(4390) and Z(4250) to be the vector tetraquark states with a relative P-wave between the diquark and antidiquark pair \cite{zgw}. To investigate the structure of the observed X(5568) state as a $0^+$  tetraquark state, Jian-Rong Zhang, Jing-Lan Zou and Jin-Yun Wu have used QCD sum rules. For this purpose, they have used four different interpolating currents, i.e. the scalar-scalar, the pseudoscalar-pseudoscalar, the axial-axial, and the vector-vector diquark-diantiquark configurations \cite{jrz}.

The prime goal to construct a quark model is to describe and estimate decay properties of the X Y Z states.  Here in the present study, we mainly focus on the charm tetraquarks having configuration $cq\bar{c}\bar{q}$, where q represents any of the light flavour u,d,s quarks. The relativistic Dirac formalism is employed to compute the confined masses of the quarks and antiquarks in a mean-field approach. Further, to calculate the masses of diquark-diantiquark system, we have numerically solved the Sch$\ddot{o}$dinger equation.  A detailed description of the theoretical formalism adopted here to study the masses of diquarks, diantiquarks  and tetraquark states with the inclusion of the spin-dependent confined one gluon exchange interactions is presented in section 2. In the third section, we present the calculations of various decay widths of the tetraquark states. In section four, we present the main results of the study and discuss keeping in view of the experimental candidates for the exotic mesonic states in the hidden charm sector.

\section{Theoretical frame work}

The theoretical methods adopted here involves two distinct parts. The first part involves computations of the inertial mass of the quarks and antiquarks by incorporating the confinement effect dynamically. These confinement masses of the quarks and antiquarks are computed using the relativistic Dirac formalism with a mean-field linear confinement potential. Using these confined quarks and antiquarks, the diquark and diantiquark states are constructed. The second part involves the computation of the binding energy of tetraquark states using diquark and diantiquark units by solving the Sch$\ddot{o}$dinger equation with an appropriate interaction potential.

 Thus, the tetraquark state here is treated as three two-body interactive systems, c-q, $\bar{c}$-$\bar{q}$ and diquark-diantiquark.  The diquark - diantiquark picture for tetraquark configuration is one of the most promising approach to understand the structure of many exotic mesonic states as the diquark-diantiquark structure is a strongly correlated system \cite{bc}. The confinement masses of quarks and antiquarks are obtained  through a mean-field linear potential of the form \cite{epjc},

\begin{equation}\label{eq:a}
V(r)= \frac{1}{2} (1+\gamma_0) (\lambda r + V_0)
\end{equation}

Where,
$\lambda$ and $V_0$ are the confinement parameters \cite{epjc,epj,prd2014,prd2016}.\\

 The single particle  wave function $\psi_q (\vec{r})$ satisfies the Dirac equation given by \cite{aruldhas,greiner},

\begin{equation}\label{eq:b}
(\vec{\alpha}\cdot\vec{p} + m_q) \psi_q (\vec{r}) = \left[ E_q - V (r) \right] \psi_q (\vec{r}),
\end{equation}

where
\begin{equation}\label{eq:d}
\alpha = \left(
    \begin{array}{cc}
     0      & \sigma \\
     \sigma & 0
    \end{array}
  \right); \qquad
\\  \gamma^0 =  \left(
    \begin{array}{cc}
     1      & 0\\
     0      & -1
    \end{array}
  \right); \qquad
\\ \gamma^i = \left(
    \begin{array}{cc}
     0      & \sigma_i \\
     -\sigma_i & 0
    \end{array}
  \right)
\end{equation}
The solution of the Dirac equation written in the two component form is expressed as \cite{greiner,epj,prd2014,prd2016},\\

\begin{equation}\label{eq:d}
\psi_{nlj}(r) = \left(
    \begin{array}{c}
      \psi_{nlj}^{(+)} \\
      \psi_{nlj}^{(-)}
    \end{array}
  \right)
\end{equation}
where
\begin{equation}\label{eq:e}
\psi_{nlj}^{(+)}(\vec{r}) = N_{nlj} \left(
    \begin{array}{c}
      i g(r)/r \\
      (\sigma.\hat{r}) f(r)/r
    \end{array}
  \right) {\cal{Y}}_{ljm}(\hat{r})
\end{equation}
\begin{equation}\label{eq:f}
\psi_{nlj}^{(-)}(\vec{r}) = N_{nlj} \left(
    \begin{array}{c}
      i (\sigma.\hat{r}) f(r)/r \\
        g(r)/r
    \end{array}
  \right) (-1)^{j+m_j-l} {\cal{Y}}_{ljm}(\hat{r})
\end{equation}

$N_{nlj}$ is the normalization constant \cite{greiner,epj,prd2014,prd2016} and the spin angular part ${\cal{Y}}_{ljm}$ is expressed as,

\begin{equation}\label{eq:g}
{\cal{Y}}_{ljm}(\hat{r}) = \sum_{m_l, m_s}\langle l, m_l, \frac{1}{2}, m_s| j, m_j \rangle Y^{m_l}_l \chi^{m_s}_{\frac{1}{2}}
\end{equation}

Here, the spinor $\chi_{\frac{1}{2}{m_s}}$ are eigenfunctions of the spin operators \cite{greiner,epj,prd2014,prd2016},
\begin{equation}\label{eq:h}
\chi_{\frac{1}{2} \frac{1}{2}} = \left(
    \begin{array}{c}
      1 \\
      0
    \end{array}
  \right)  = |\alpha> \ \ \ , \ \ \ \ \chi_{\frac{1}{2} -\frac{1}{2}} = \left(
    \begin{array}{c}
      0 \\
      1
    \end{array}
  \right)= |\beta>
\end{equation}

The reduced radial part $g(r)$ and $f(r)$ of the Dirac spinor $\psi_{nlj}(r)$ are the solutions of the equations given by \cite{greiner,epj,prd2014,prd2016},\\

\begin{equation}\label{eq:i}
\frac{d^2 g(r)}{dr^2}+\left[(E_{D}+ m_q)[E_{D} - m_q - V(r)]-\frac{\kappa(\kappa + 1)}{r^2}\right] g(r) = 0
\end{equation}
and
\begin{equation}\label{eq:j}
\frac{d^2 f(r)}{dr^2}+\left[(E_{D}+ m_q)[E_{D}- m_q - V(r)]-\frac{\kappa(\kappa-1)}{r^2}\right] f(r) = 0
\end{equation}

it is good approximation to define a new quantum number $\kappa$ \cite{greiner,epj,prd2014,prd2016} as,
\begin{equation}\label{eq:k}
  \kappa = \left \{ \begin{array}{c}
      -(l+1)       \ \ \  for \ \ \   j = l + \frac{1}{2}\\\\

      l          \ \ \   for \ \ \  j = l - \frac{1}{2}
    \end{array}\right.
\end{equation}

Eq. (10) and (11) can be expressed in dimensionless form as, \cite{aruldhas,epj,prd2014,prd2016},
\begin{equation}\label{eq:l}
\frac{d^2 g(\rho)}{d\rho^2}+\left[\epsilon - \rho^{1.0} - \frac{\kappa (\kappa + 1)}{\rho^2}\right] g(\rho) = 0
\end{equation}

\begin{equation}\label{eq:m}
\frac{d^2 f(\rho)}{d\rho^2}+\left[ \epsilon - \rho^{1.0} - \frac{\kappa (\kappa - 1)}{\rho^2}\right] f(\rho) = 0
\end{equation}

Where, $\rho= \frac{r}{r_0}$ is a dimensionless variable with suitably chosen scale  factor
$r_0 =\frac{r}{[(E+m)\lambda] ^\frac{-1}{3}}$ and $\epsilon$ is expressed as \cite{epj,prd2014,prd2016},

\begin{equation}\label{eq:n}
\epsilon = (E_D-m_q-V_0)(m_q+E_D)^\frac{1}{3} \lambda^\frac{-2}{3}
\end{equation}

The solution of $f(\rho)$ and $g(\rho)$ are normalized to get,

\begin{equation}\label{eq:o}
\int^\infty_0 \left[f^2(\rho)+g^2(\rho)\right]d\rho = 1
\end{equation}

The wavefunction for two body system $\Psi(1,2)$ can be constructed by combining two single-particle wave functions \cite{vaghmare} :

\begin{eqnarray}
\Psi(1,2)  &=& \frac{1}{\sqrt{2}} [\psi_{n_1,l_1,j_1} (r_1) \psi_{n_2,l_2,j_2} (r_2) - \\&& \nonumber  \psi_{n_2,l_2,j_2} (r_2) \psi_{n_1,l_1,j_1} (r_1)] \chi^{AS}_{12}
\end{eqnarray}

\begin{eqnarray}
\Psi(1,2)  &=& \frac{1}{\sqrt{2}} [\psi_{n_1,l_1,j_1} (r_1) \psi_{n_2,l_2,j_2} (r_2) + \\&& \nonumber  \psi_{n_2,l_2,j_2} (r_2) \psi_{n_1,l_1,j_1} (r_1)] \chi^{S}_{12}
\end{eqnarray}

corresponds to spin anti-symmetric $\vec{J}$ = $\vec{0}$ state and spin symmetric $\vec{J}$ = $\vec{1}$ state respectively.\cite{vaghmare}. The  spin anti-symmetric state is given by,

Here,
\begin{equation}
\chi^A_{12} = \frac{1}{\sqrt{2}}(\alpha(1) \beta(2) - \alpha(2) \beta(1))
\end{equation}

And the three states with spin  $\vec{J}$ = $\vec{1}$ with $m_J$ = 1, 0, -1  triplet configuration expressed as \cite{vaghmare},

\begin{equation}
\chi^S_{12}(J = 1, m_J = 1) = \alpha(1) \alpha(2)
\end{equation}

\begin{equation}
\chi^S_{12}(J = 1, m_J = 0) = \frac{1}{\sqrt{2}}(\alpha(1) \beta(2) + \alpha(2) \beta(1)))
\end{equation}

\begin{equation}
\chi^S_{12}(J = 1, m_J = -1) = \beta(1) \beta(2)
\end{equation}

The confined effective mass of a two body system of our interest can be obtained as \cite{epj,prd2014,prd2016},

\begin{equation}
M^{eff}_{12} = (E_{1} + m_{1}) + (E_{2} + m_{2}) - E^{cm}_{12}
\end{equation}

Where, $E_{1/2}$ represents the Dirac's single-particle energies and $E^{cm}_{12}$ represents the centre of mass of the two-particle system. Here, we treat  $E^{cm}_{12}$  as a parameter and absorb with the potential parameter $V_0$.\\

\begin{table*}
\begin{center}
\tabcolsep4.0pt
   \small
\caption{Fitted Model Parameters} \label{tab1}
\begin{tabular}{c |c c c}
\hline
Quark  &    Systems  &    potential  & $V_0$ \\
 Masses  &     &    strength &  \\
 ($GeV/c^2$)      &     &     $(GeV^2)$ & GeV  \\
\hline
$m_c$ = 1.27 &   $c\bar{q}$/$cq$/$\bar{c}\bar{q}$ & $\lambda_{cq}$ = 0.04 & -C.F$\ast$$\alpha_s$$\ast$$\frac{0.35}{(n+l+1)^{1.5}}$\\
$ m_q $ = 0.3  & $c\bar{s}$/$cs$/$\bar{c}\bar{s}$ &  $\lambda_{cs}$ = 0.048 & -C.F $\ast$$\alpha_s$$\ast$$\frac{0.41}{(n+l+1)^{1.5}}$\\
$ m_s $= 0.5  &  $cq\bar{c}\bar{q}$   & 4$\ast$($\lambda_{cq}$)  &  -C.F $\ast$$\alpha_s$$\ast$$\frac{0.35}{(n+l+1)^{1.5}}$ \\
&  $cs\bar{c}\bar{s}$  & 4$\ast$($\lambda_{cs}$)  & -C.F $\ast$$\alpha_s$$\ast$$\frac{0.41}{(n+l+1)^{1.5}}$\\
&  $cs\bar{c}\bar{q}$  & 4$\ast$($\frac{\lambda_{cq} + \lambda_{cs} }{2}$) & $\frac{ V_0(cq) + V_0(cs)}{2}$ \\
\hline
\multicolumn{4}{l}{C.F = Colour Factor, $\alpha_s$ = Strong running coupling constant}
\end{tabular}
\end{center}
\end{table*}

\subsection{\textbf{Masses of Diquark / Diantiquarks and tetraquark system}}

We assume the formation of the diquark is due to the linear and the confined one gluon exchange potential. For the quark - quark interaction in a diquark, the potential between the two quarks qq/$\bar{q}$$\bar{q}$ is half the one between a quark q and antiquark $\bar{q}$, i.e  $V_{qq}$ = 1/2 $V_{q\bar{q}}$. The confined quark/antiquark masses are employed  to obtain the spin average masses of  diquark and diantiquark states. In the charm sector, we compute the masses of the diquarks ($M_d$) and diantiquarks ($M_{\bar{d}}$) as,

\begin{equation}
M_d = (E_{c}+m_{c}) + (E_{q}+m_{q}) - E^{cm}({cq})
\end{equation}

\begin{equation}
M_{\bar{d}} = (E_{\bar{c}}+ m_{\bar{c}}) + (E_{\bar{q}}+ m_{\bar{q}})- E^{cm}({\bar{c} \bar{q}})
\end{equation}

Here, d and $\bar{d}$ represent diquark and diantiquark, respectively. Further the spin dependent part of the c-q and $\bar{c}$-$\bar{q}$ interactions are considered perturbingly. Accordingly the masses of cq, $\bar{c}\bar{q}$ states are computed as,

\begin{equation}
M_{J_d} = M_d + \langle V^{j_1 j_2}_{cq} \rangle + \langle V^{L S}_{cq} \rangle + \langle V^{T}_{cq} \rangle
\end{equation}

\begin{equation}
M_{J_{\bar{d}}} = M_{\bar{d}} + \langle V^{j_1 j_2}_{{\bar{c}}{\bar{q}}} \rangle + \langle V^{L S}_{{\bar{c}}{\bar{q}}} \rangle + \langle V^{T}_{{\bar{c}}{\bar{q}}} \rangle
\end{equation}

Where,  $\langle V^{j_1 j_2} \rangle$, $\langle V^{L S} \rangle$ and $\langle V^{T} \rangle$ are the Spin - Spin, Spin - Orbit and Tensor interactions.  The model parameters are fixed by constructing the conventional mesonic (q$\bar{q}$) state where, $V_{qq}$ = 1/2 $V_{q\bar{q}}$  has been considered. The potential parameters including masses of quark and potential strength  are chosen in such a way that mass of tetra quark states should satisfies the mass inequality relation $X_{cq\bar{c}\bar{q}}$ $\leq$  $ 2 M_{c\bar{q}}(1s)$ \cite{mna}.
The computed masses of the $c\bar{q}$, cq and $\bar{c}\bar{q}$ states with  q $\epsilon$ u,d,s states are listed in Table  \ref{tab2} for the low lying  S-wave and P-waves.
Now, using the diquark (qq) and diantiquark ($\bar{q}$$\bar{q}$) states, the tetra quark system has been studied non relativistically with the similar linear interaction potential assumed between the diquark-diantiquark. Accordingly, the mass of tetraquark state is expressed as,

\begin{equation}
M_{d\bar{d}} = M_{d} + M_{\bar{d}} + E_{d\bar{d}} +V_{SD}({d\bar{d}})
\end{equation}

Where, the binding energy $E_{d\bar{d}}$ is obtained by numerically solving the Schr$\ddot{o}$dinger equation using linear plus constant potential. And the $V_{SD}(d\bar{d})$ is the spin dependent interaction among the diquark-diantiquark system.  Finally, the mass of the tetraquark system is expressed as,

\begin{equation}
M_{J_d J_{\bar{d}}}(d\bar{d})= M_{d\bar{d}} + \langle V^{j_d j_{\bar{d}}}_{d\bar{d}} \rangle + \langle V^{L S}_{d\bar{d}} \rangle + \langle V^{T}_{d\bar{d}} \rangle
\end{equation}

\subsection{\textbf{Spin dependent two body interactions}}

Although the simple model described above seems sufficient for computing the mass of two-body systems, the full mass spectrum can be obtained by incorporating the spin-dependent part of the interactions. The spin-spin, spin-orbit and tensor interactions of confined one gluon exchange potentials (COGEP) \cite{epj,prd2014,prd2016} are employed to remove the degeneracy of the two-body systems under study.\\

The mass of the two body system represented by $M_{^{2s+1}L_J}$ \cite{epj,prd2014,prd2016} thus becomes,\\
\begin{eqnarray}\label{eq:q}
M_{^{2s+1}L_J} &=& M_{12}(n_1l_1j_1,n_2l_2j_2)+ \langle V^{j_1 j_2}_{{12}} \rangle+
\langle V^{LS}_{12} \rangle +\nonumber \\&& \langle V^{T}_{12} \rangle
\end{eqnarray}

Considering the problem of two body system, two independent angular momenta are added. Thus, for a two particle system consisting of two with angular momenta $J_1$ and $J_2$ couple to the total angular momentum J as \cite{bsrajput},

\begin{equation}
\overrightarrow{J} = \overrightarrow{J_1} + \overrightarrow{J_2}
\end{equation}
The commutation relations obeyed by $J_1$ and $J_2$ \cite{bsrajput},

\begin{equation}
[\hat{J}_{1i},\hat{J}_{1j}] = i\hbar\epsilon_{ijk}\hat{J}_{1k}
\end{equation}

\begin{equation}
[\hat{J}_{2i},\hat{J}_{2j}] = i\hbar\epsilon_{ijk}\hat{J}_{2k}
\end{equation}

\begin{equation}
[\hat{J}_{1i},\hat{J}_{2j}] = 0
\end{equation}

Here, i, j and k running from 1 to 3, denoting x, y and z directions, respectively and  $\epsilon$ is the Levi-Civita tensor.

The eigenstates of $J_1^2$ and $J_{1z}$ ,  $|j_1,m_1\rangle$ and that of  $J_2^2$ and $J_{2z}$ ,  $|j_2,m_2\rangle$ are written as,

\begin{equation}
|j_1j_2m_1m_2\rangle \equiv |j_1,m_1\rangle \otimes |j_2,m_2\rangle
\end{equation}

For basis set(1) $J_1^2$, $J_2^2$, $J_{1z}$ and $J_{2z}$ the eigenstate is designated as $|j_1j_2m_1m_2\rangle$ . The quantum numbers are obtained as \cite{bsrajput},

\begin{equation}
J_1^2|j_1j_2m_1m_2\rangle  = j_1(j_1+1) \hbar^2 |j_1j_2m_1m_2\rangle
\end{equation}

\begin{equation}
J_{1z}|j_1j_2m_1m_2\rangle  = m_1\hbar |j_1j_2m_1m_2\rangle
\end{equation}

\begin{equation}
J_2^2|j_1j_2m_1m_2\rangle  = j_2(j_2+1) \hbar^2 |j_1j_2m_1m_2\rangle
\end{equation}

\begin{equation}
J_{2z}|j_1j_2m_1m_2\rangle  = m_2\hbar |j_1j_2m_1m_2\rangle
\end{equation}

For basis set (2) the eigenstate is designated as $|j_1j_2jm\rangle$. Now we have \cite{bsrajput},

\begin{equation}
J^2|j_1j_2jm\rangle  = j(j+1) \hbar^2 |j_1j_2m_1m_2\rangle
\end{equation}
\begin{equation}
J_1^2|j_1j_2jm\rangle  = j_1(j_1+1) \hbar^2 |j_1j_2jm\rangle
\end{equation}

\begin{equation}
J_2^2|j_1j_2jm\rangle  = j_2(j_2+1) \hbar^2 |j_1j_2jm\rangle
\end{equation}

\begin{equation}
J_z|j_1j_2jm\rangle  = m\hbar |j_1j_2jm\rangle
\end{equation}

Now

\begin{equation}
[J^2,J_{1z}] \neq 0
\end{equation}

\begin{equation}
[J^2,J_{2z}] \neq 0
\end{equation}

Equation (41) to (48) show that basis set (1) and basis set (2) consist different bases in the same Hilbert space. The solution of this problem is that we have to use linear combination of $|j_1j_2m_1m_2\rangle$ states and $|j_1j_2jm\rangle$ states vice-versa \cite{bsrajput}. Now,

\begin{equation}
\sum_{j_1j_2} \sum_{m_1m_2}|j_1j_2m_1m_2\rangle \langle j_1j_2m_1m_2| = 1
\end{equation}

\begin{equation}
\sum_{j_1j_2} \sum_{jm}|j_1j_2jm\rangle \langle j_1j_2jm| = 1
\end{equation}

Using unitary transformation one can go fron set (1) to set (2) through the C.G coefficient \cite{bsrajput},

\begin{equation}
|j_1j_2jm\rangle = \sum_{m_1m_2}  \langle j_1j_2m_1m_2|\langle j_1j_2m_1m_2 | j_1j_2jm\rangle
\end{equation}

The transformation matrix $C_{jmm_1m_2} = \langle j_1j_2m_1m_2|j_1j_2jm \rangle$ is unitary. The elements  $\langle j_1j_2m_1m_2 | j_1j_2jm\rangle$  sare called Clebsch-Gordan
coefficients \cite{bsrajput}.

For spin-spin interaction, the j-j coupling term is expressed as \cite{epj,amp1,prd2014,prd2016},\\

 \begin{equation}\label{eq:r}
\langle V^{j_1 j_2}_{12} \rangle = \frac{\sigma\langle j_1 j_2 j m|\widehat{j_1}\widehat{j_2}| j_1 j_2 j m\rangle}
{(E_1 + m_1)(E_{2} + m_{2})}
  \end{equation}

$\langle j_1 j_2 j m|\widehat{j_1}\widehat{j_2}| j_1 j_2 j m\rangle $ contains the
  square of the Clebsch- Gordan coefficient. The spin orbit interaction and tensor
  interactions are expressed respectively as  \cite{epj,prd2014,prd2016,amp1},\\

\begin{eqnarray}\label{eq:s}
\langle V^{L S}_{12} \rangle &=& \frac{\alpha_s}{4} \frac{N_1^2 N_{2}^2}
{(E_1 + m_1)(E_{2} + m_{2})}
\frac{\lambda_1\lambda_{2}}{2r} \\ &&
\otimes[\overrightarrow {r} \times (\widehat{p}_1-\widehat{p}_{2})
.(\sigma_1-\sigma_{2})] (D_0'(r)+2 D_1'(r)) \nonumber \\&&
+ [ [\overrightarrow {r} \times (\widehat{p}_1 + \widehat{p}_{2}).
(\sigma_i-\sigma_j)(D_0'(r) - D_1'(r)) \nonumber
\end{eqnarray}

and

\begin{eqnarray}\label{eq:t}
\langle V^{T}_{{12}} \rangle &=& -\frac{\alpha_s}{4} \frac{N_1^2 N_{2}^2}
{(E_1 + m_1)(E_{2} + m_{2})}
\lambda_1\lambda_{2} \\ &&
\otimes ((\frac{D_1''(r)}{3}-\frac{D_1'(r)}{3r}) S_{1{2}}) \nonumber
\end{eqnarray}

where, $S_{12} = \left[ 3 (\sigma_1. {\hat{r}})(\sigma_{2}.
{\hat{r}})- \sigma_1 . \sigma_{2}\right]$ and ${\hat{r}} = {\hat{r}}_1 -
{\hat{r}}_{2}$ is the unit vector in the relative coordinate  \cite{epj,prd2014,prd2016}.\\

The running strong coupling constant $\alpha_s$ is computed as \cite{ebert},\\

\begin{equation}
 \alpha_s = \frac{4 \pi}{(11-\frac{2}{3}\  n_\emph{f})\log\left(\frac{\mu^2 + M^2_B}{\Lambda^2_{QCD}}\right)}
\end{equation}

with $n_\emph{f}$ = 3, the background mass $M_B$ = $0.95$  GeV \cite{ebert} and $\Lambda_{QCD}$ = 0.250 GeV. We have adopted the  form of the confined gluon propagators which are given by \cite{pcv,amp1,epj,prd2014,prd2016},\\
\begin{equation}
D_0 (r) = \left( \frac{\alpha_1}{r}+\alpha_2 \right) \exp(-r^2 c_0^2/2)
\end{equation}
and
\begin{equation}
D_1 (r) =  \frac{\gamma}{r} \exp(-r^2 c_1^2/2)
\end{equation}

where $\alpha_1$ = 1.035, $\alpha_2$ = 0.3977, $c_0$ = 0.3418 GeV , $c_1$ = 0.4123 GeV,  $\gamma$ = 0.8639 are the fitted parameter as in  \cite{amp1}.  The basic parameters of the present model are fixed to yield the expression of the masses of the $Q\bar{q}$ ( D, $D_s$) mesonic spectra. The resulting parameters are listed in Table \ref{tab1}.  The computed results for the masses are listed in Table \ref{tab3} to Table \ref{tab6}.

\begin{table*}
\begin{center}
\tabcolsep4.0pt
   \small
\caption{S wave and P wave masses for $Q\bar{q}$ (Q $\epsilon$ c and q $\epsilon$ u,d,s)  mesons and diqaurks (Qq) / diantiquarks $(\bar{Q}\bar{q})$ (in MeV)} \label{tab2}
\begin{tabular}{c c c c c c c c c}
\hline\hline
state    &    State     &      $c\bar{s} $    &    $c\bar{s} $    &    $cs/\bar{c}\bar{s}$    &    $c\bar{q} $    &    $c\bar{q} $    &    $cq/\bar{c}\bar{q}$    \\
    &    notation    &            (Our)    &    (Exp.)    &        &    (Our)    &    (Exp.)    &        \\
\hline\hline
1S    &    $^3S_1$    &       2110    &    $2112.2  \pm  0.4$     &    2332    &    2010    &    $2010.2 \pm 0.05$    &    2207    \\
    &    $^1S_0$    &       1973&    $1968.34 \pm 0.07$    &    2208    &    1861    &    $1869.65 \pm 0.05$    &    2075    \\
\\                                                                    \\
2S    &    $^3S_1$    &        2728    &    $2708.3^{+4.0}_{-3.4}$    &    2805    &    2609    &    2608.7\cite{mns}    &    2674\\
    &    $^1S_0$    &     2630    &    2638\cite{mns}    &    2714&    2503    &    2539.4\cite{mns}    &    2579    \\
\\                                                                    \\
1P    &    $^3P_2$    &        2543    &    $2571.9 \pm 0.8$    &    2646    &    2491    &    $2426.6 \pm 0.7$    &    2549    \\
    &    $^3P_1$    &       2425    &    $2459.6 \pm 0.6$    &    2584    &    2342    &    $\cdots$    &    2471    \\
    &    $^3P_0$    &        2335    &    $2317.8 \pm 0.6$    &    2534    &    2318    &    $2318 \pm 29$    &    2455    \\
    &    $^1P_1$    &       2526    &    $2535.12 \pm 0.13$    &    2589    &    2400&    $2421.3 \pm 0.6$    &    2453    \\

\hline
\end{tabular}
\end{center}
\end{table*}

\begin{table*}
\begin{center}
\tabcolsep4.0pt
   \small
\caption{Mass spectra of $cs\bar{c}\bar{s}$, $cq\bar{c}\bar{q}$ and $cs\bar{c}\bar{q}$ states in the diquark $-$ diantiquark picture for $L_d = 0$ and $L_{\bar{d}} = 0$ (in MeV)} \label{tab3}
\begin{tabular}{c c c c c c c c c cccc}
\hline\hline
state    &    $S_d$    &    $L_d$    &    $S_{\bar{d}}$    &    $L_{\bar{d}}$    &    $J_d$    &    $J_{\bar{d}}$    &    $J =$    &$J^{PC}$&state&         & Masses of \\
    &        &    &        &        &        &        & $J_d+J_{\bar{d}}$    &&notation&     $cs\bar{c}\bar{s}$    & $cq\bar{c}\bar{q}$ &  $cs\bar{c}\bar{q}$\\
\hline\hline
1s    &    0    &    0    &    0    &    0    &    0    &    0    &    0    &    $0^{++}$    &    $^1S_0$    &    3967    &    3739&    3852    \\
\\
    &    1    &    0    &    0    &    0    &    1    &    0    &    1    &    $1^{+\pm}$    &    $^3S_1$    &    4097    &    3877&    3981\\
\\
    &    1    &    0    &    1    &    0    &    1    &    1    &    0    &    $0^{++}$    &    $^1S_0$    &    4214    &    4001    &    4106    \\
    &        &        &        &        &        &        &    1    &    $1^{+-}$    &    $^3S_1$    &    4217    &    4004    &    4110    \\
    &        &        &        &        &        &        &    2    &    $2^{++}$    &    $^5S_2$    &    4229    &    4018    &    4123    \\
    &        &        &        &        &        &        &        &        &        &        &        &        \\
2s    &    0    &    0    &    0    &    0    &    0    &    0    &    0    &    $0^{++}$    &    $^1S_0$    &    4505    &    4325    &    4414    \\
\\
    &    1    &    0    &    0    &    0    &    1    &    0    &    1    &    $1^{+\pm}$    &    $^3S_1$    &    4601    &    4425    &    4510    \\
\\
    &    1    &    0    &    1    &    0    &    1    &    1    &    0    &    $0^{++}$    &    $^1S_0$    &    4689    &    4516    &    4601    \\
    &        &        &        &        &        &        &    1    &    $1^{+-}$    &    $^3S_1$    &    4691    &    4518    &    4604    \\
    &        &        &        &        &        &        &    2    &    $2^{++}$    &    $^5S_2$    &    4700    &    4528    &    4612    \\
\hline
\end{tabular}
\end{center}
\end{table*}

\begin{table*}
\begin{center}
\tabcolsep4.0pt
   \small
\caption{Mass spectra of $cs\bar{c}\bar{s}$ and $cq\bar{c}\bar{q}$ states in the diquark - diantiquark picture for $L_d = 1$ and $L_{\bar{d}} = 0$ (in MeV)} \label{tab4}
\begin{tabular}{c c c c c c c c ccccc}
\hline\hline
State    &    $S_d$    &    $L_d$    &    $S_{\bar{d}}$    &    $L_{\bar{d}}$    &    $J_d$    &    $J_{\bar{d}}$    &    $J=$    &    $J^{PC}$    &    state    &        &    Masses of     \\
    &        &        &        &        &        &        &    $J_d+J_{\bar{d}}$    &        &    notation    &    $cs\bar{c}\bar{s}$    &    $cq\bar{c}\bar{q}$    & $cs\bar{c}\bar{q}$    \\
\hline\hline
1P    &    0    &    1    &    0    &    0    &    1    &    0    &    1    &    $1^{--}$    &    $^1P_1$    &    4416    &    4217&    4315    \\
\\
    &    1    &    1    &    0    &    0    &    0    &    0    &    0    &    $0{-+}$    &    $^3P_0$    &    4428    &    4269    &    4328    \\
    &        &        &        &        &    1    &        &    1    &    $1{-+}$    &    $^3P_1$    &    4439    &    4278    &    4339    \\
    &        &        &        &        &    2    &        &    2    &    $2{-+}$    &    $^3P_2$    &    4445    &    4283    &    4344    \\
\\
    &    1    &    1    &    1    &    0    &    0    &    1    &    1    &    $1^{--}$    &    $^1P_1$    &    4466&    4343    &    4404    \\
    &        &        &        &        &    1    &    1    &    0    &    $0{-+}$    &    $^3P_0$    &    4385    &    4296    &    4339    \\
    &        &        &        &        &        &        &    1    &    $1{-+}$    &    $^3P_1$    &    4413    &    4314    &    4362\\
    &        &        &        &        &        &        &    2    &    $2{-+}$    &    $^3P_2$    &    4487    &    4355&    4420    \\
    &        &        &        &        &    2    &    1    &    1    &    $1{--}$    &    $^5P_1$    &    4414    &    4308&    4359    \\
    &        &        &        &        &        &        &    2    &    $2{--}$    &    $^5P_2$    &    4420    &    4320    &    4369    \\
    &        &        &        &        &        &        &    3    &    $3{--}$    &    $^5P_3$    &    4555    &    4392    &    4474    \\
\hline
\end{tabular}
\end{center}
\end{table*}

\begin{table*}
\begin{center}
\tabcolsep4.0pt
   \small
\caption{Comparison of predicted masses of $cq\bar{c}\bar{q}$ with the available theoretical approaches (in MeV)} \label{tab5}
\begin{tabular}{c c c c c c c c cccc}
\hline\hline
state    &    $S_d$    &    $L_d$    &    $S_{\bar{d}}$    &    $L_{\bar{d}}$    &    $J_d$    &    $J_{\bar{d}}$    &    Total  J    &    Our     &    \cite{ebert2008}    &    \cite{epja2014}    \\
    &        &        &        &        &        &        &        &        &        &        \\
\hline\hline
1s    &    0    &    0    &    0    &    0    &    0    &    0    &    0    &    3739    &    3812    &    3906    \\
    &    1    &    0    &    0    &    0    &    1    &    0    &    1    &    3877    &    3871    &    3910    \\
    &    1    &    0    &    1    &    0    &    1    &    1    &    0    &    4001    &    3852    &    3849    \\
    &        &        &        &        &        &        &    1    &    4004    &    3891    &    3882    \\
    &        &        &        &        &        &        &    2    &    4018    &    3968    &    3946    \\
                                                                                    \\
2s    &    0    &    0    &    0    &    0    &    0    &    0    &    0    &    4325    &    $\cdots$    &    $\cdots$    \\
    &    1    &    0    &    0    &    0    &    1    &    0    &    1    &    4425&    $\cdots$    &    $\cdots$    \\
    &    1    &    0    &    1    &    0    &    1    &    1    &    0    &    4516    &    $\cdots$    &    $\cdots$    \\
    &        &        &        &        &        &        &    1    &    4518    &    $\cdots$    &    $\cdots$    \\
    &        &        &        &        &        &        &    2    &    4528    &    $\cdots$    &    $\cdots$    \\
\\
1P    &    0    &    1    &    0    &    0    &    1    &    0    &    1    &    4217    &    4244    &    4164    \\
    &    1    &    1    &    0    &    0    &    0    &    0    &    0    &    4269    &    $\cdots$    &    4136    \\
    &        &        &        &        &    1    &        &    1    &    4278    &    $\cdots$    &    4159    \\
    &        &        &        &        &    2    &        &    2    &    4283    &    $\cdots$    &    4154    \\
    &    1    &    1    &    1    &    0    &    0    &    1    &    1    &    4343    &    $\cdots$    &    4145    \\
    &        &        &        &        &    1    &    1    &    0    &    4296    &    $\cdots$    &    4128    \\
    &        &        &        &        &        &        &    1    &    4314    &    $\cdots$    &    4151    \\
    &        &        &        &        &        &        &    2    &    4355    &    $\cdots$    &    4146    \\
    &        &        &        &        &    2    &    1    &    1    &    4308&    $\cdots$    &    4113    \\
    &        &        &        &        &        &        &    2    &    4320    &    $\cdots$    &    4174    \\
    &        &        &        &        &        &        &    3    &    4392&    $\cdots$    &    4142    \\
\hline
\end{tabular}
\end{center}
\end{table*}

\begin{table*}
\begin{center}
\tabcolsep3.0pt
   \small
\caption{Comparison of predicted masses of $cs\bar{c}\bar{s}$ with the available theoretical approaches (in MeV)} \label{tab6}
\begin{tabular}{c c c c c c c c ccc}
\hline\hline
state    &    $S_d$    &    $L_d$    &    $S_{\bar{d}}$    &    $L_{\bar{d}}$    &    $J_d$    &    $J_{\bar{d}}$    &    Total  J    &    Our     &    \cite{ebert2008}    \\
\hline\hline
1s    &    0    &    0    &    0    &    0    &    0    &    0    &    0    &    3967    &    4051    \\
    &    1    &    0    &    0    &    0    &    1    &    0    &    1    &    4097    &    4113    \\
    &    1    &    0    &    1    &    0    &    1    &    1    &    0    &    4214    &    4110    \\
    &        &        &        &        &        &        &    1    &    4217    &    4143    \\
    &        &        &        &        &        &        &    2    &    4229    &    4209    \\
                                                                            \\
2s    &    0    &    0    &    0    &    0    &    0    &    0    &    0    &    4505&    $\cdots$    \\
    &    1    &    0    &    0    &    0    &    1    &    0    &    1    &    4601    &    $\cdots$    \\
    &    1    &    0    &    1    &    0    &    1    &    1    &    0    &    4689    &    $\cdots$    \\
    &        &        &        &        &        &        &    1    &    4691    &    $\cdots$    \\
    &        &        &        &        &        &        &    2    &    4700&    $\cdots$    \\
                                                                            \\
1P    &    0    &    1    &    0    &    0    &    1    &    0    &    1    &    4416    &    4466    \\
    &    1    &    1    &    0    &    0    &    0    &    0    &    0    &    4428    &    $\cdots$    \\
    &        &        &        &        &    1    &        &    1    &    4439    &    $\cdots$    \\
    &        &        &        &        &    2    &        &    2    &    4445    &    $\cdots$    \\
    &    1    &    1    &    1    &    0    &    0    &    1    &    1    &    4466    &    $\cdots$    \\
    &        &        &        &        &    1    &    1    &    0    &    4385    &    $\cdots$    \\
    &        &        &        &        &        &        &    1    &    4413&    $\cdots$    \\
    &        &        &        &        &        &        &    2    &    4487    &    $\cdots$    \\
    &        &        &        &        &    2    &    1    &    1    &    4414    &    $\cdots$    \\
    &        &        &        &        &        &        &    2    &    4420    &    $\cdots$    \\
    &        &        &        &        &        &        &    3    &    4555&    $\cdots$    \\
\hline
\end{tabular}
\end{center}
\end{table*}

\begin{table*}
\begin{center}
\tabcolsep1.2pt
   \small
\caption{Radiative and leptonic decay widths of $1^{--}$ states} \label{tab7}
\begin{tabular}{c| c |c| c |c |c |c}
\hline\hline
Experimental    &    Experimental     &    Predicted Mass    &    Quark composition    &    Decay  modes    &    Predicted    &    Experimental     \\
state    &    Mass  (MeV)    &    (MeV)    &        &        &    Decay width    &    Decay width    \\
\hline
$\psi(4230)$    &    $\sim 4230$    &    4217    &    $cq\bar{c}\bar{q}$    &    $\Gamma_{(\psi(4230 \rightarrow J/\psi \gamma)}$    &    1.932 MeV    &      $\cdots$    \\
    &        &        &        &    $\Gamma_{e^+e^-}$    &    0.391 keV    &      $\cdots$    \\
    &        &        &        &        &        &        \\
$\psi(4260)$    &    4230$ \pm$ 8    &    4263    &    $cq\bar{c}\bar{q}$    &    $\Gamma_{(\psi(4260 \rightarrow J/\psi \gamma)}$    &    1.959 MeV    &      $\cdots$    \\
    &        &        &        &    $\Gamma_{e^+e^-}$    &    0.374 keV    &      $\cdots$    \\
    &        &        &        &        &        &        \\
$\psi(4360)$    &    4359 $\pm$ 13     &    4359    &    $cs\bar{c}\bar{q}$    &    $\Gamma_{(\psi(4360 \rightarrow J/\psi \gamma)}$    &    2.009 MeV    &      $\cdots$    \\
    &        &        &        &    $\Gamma_{e^+e^-}$    &    0.369 keV    &      $\cdots$    \\
    &        &        &        &        &        &        \\
$\psi(4390)$    &    $\sim 4390$    &    4404     &    $cs\bar{c}\bar{q}$    &    $\Gamma_{(\psi(4390 \rightarrow J/\psi \gamma)}$    &    2.030 MeV    &      $\cdots$    \\
    &        &        &        &    $\Gamma_{e^+e^-}$    &    0.395 keV    &      $\cdots$    \\
    &        &        &        &        &        &        \\
$\psi(4415)$    &    4421 $\pm$ 4    &    4416 &    $cs\bar{c}\bar{s}$    &    $\Gamma_{(\psi(4415 \rightarrow J/\psi \gamma)}$    &    2.035 MeV    &      $\cdots$    \\
    &        &        &        &    $\Gamma_{e^+e^-}$    &    0.498  keV    &    0.58 $\pm$ 0.07    \\
    &        &        &        &        &        &        \\
\hline
\end{tabular}
\end{center}
\end{table*}

\section{Decay properties of Exotic Tetraquark states}
In addition to the mass spectra, predictions of the decay widths play a crucial role in the identification of the structure and quark compositions of the exotic states. The radiative decay is believed to be an ideal tool to understand the hadronic structure of newly observed resonances as these transitions are driven by the electromagnetic interactions only \cite{hwk}. Thus, in the present study, we calculate the radiative decay rates of heavy-light tetraquarks states ($cq\bar{c}\bar{q}$, q $\epsilon$ u/d,s)  particularly in the channel $X \rightarrow J/\psi \gamma$. The radiative decay of the four quark state is calculated using the concept of Vector meson dominance (VMD). VMD explains the interactions between photon and hadronic matter \cite{jjs}. The transition matrix for the decay process  $X \rightarrow J/\psi \gamma$ is written as,\\

\begin{equation}
\langle J/\psi \gamma |X \rangle = \langle \gamma|\rho\rangle \frac{1}{m_\rho^2} \langle J/\psi \rho|X \rangle = \frac{f_\rho}{m_\rho ^2}A
\end{equation}

Where A is taken as the same as used in Ref.  \cite{lm_2004,ar}.

Thus, the partial decay width is given by \cite{ar},\\
\begin{eqnarray}
\Gamma ( X \rightarrow J/\psi \gamma) &=& 2|A^2| \left(\frac{f_\rho}{m_\rho ^2}\right)^2 \frac{1}{8\pi M_X^2} \nonumber \\&&
\frac{\sqrt{\lambda(M_X, M_\psi,0)}}{2M_X}
\end{eqnarray}

Where, $f_\rho = 0.152 GeV^2$ \cite{pad,epja2014} and the decay momentum $\lambda(a,b,c)$ for the decay process $a \rightarrow bc$ is given by \cite{ar},\\

\begin{eqnarray}
\lambda &=& (M_a)^4 + (M_b)^4 + (M_c)^4 - 2 ( M_a M_b)^2 - 2 ( M_a M_c)^2 \nonumber \\&&
- 2 ( M_b M_c)^2
\end{eqnarray}

The leptonic Decay width of charm tetra quark state is computed  using Van Royen–Weisskopf formula for P-waves \cite{sjp2,ahmedali} as,

\begin{equation}
\Gamma_{e^+e^-} = \frac{24\alpha^2<e_q^2>}{M^4}\sigma^2 |R'(0)|^2
\end{equation}

The radiative decay widths of some of the $\psi$ states in the 4 to 5 GeV range and their leptonic decay widths are listed in Table \ref{tab7}. Another important decay channel for the exotic meson in the hidden charm sector is $X \rightarrow D_q\bar{D}_q$. And according to latest Particle Data Group 2018 \cite{pdg2018}, $D\bar{D}$ decay channel have been observed experimentally for some of $\psi$ state even though its decay width is not reported. In 2016, Ruilin Zhu have investigated the masses and decays of hidden charm tetra quarks \cite{zhu2016}.We have adopted the same formalism to calculate the decay width of $J^{PC}$ = $1^{--}$ states. The computed hadronic decay widths for $\psi$ states using following equations,

\begin{equation}
\Gamma (\psi (M) \rightarrow D_q \bar{D_q}) = \frac{F^2 |\overrightarrow{K}|^3}{6 M^2 \pi}
\end{equation}

\begin{equation}
\Gamma (\psi (M) \rightarrow D_q \bar{D_q}^*) = \frac{F^2 |\overrightarrow{K}|^3}{12 M^2 \pi}
\end{equation}

Where, F is effective coupling, and it is adopted from Ref \cite{rsa2004}and  q $\epsilon$ u/d, s

\begin{equation}
|\overrightarrow{K}| = \frac{\sqrt{M^2 - (M_1 + M_2)^2} \sqrt{M^2 - (M_1 + M_2)^2}}{2M}
\end{equation}

Where, $|\overrightarrow{K}|$ is the center of mass momentum, $M_1$ and $M_2$ are the masses of the decay products \cite{sjp2} and M is the mass of decaying particle \cite{sjp2}.  The computed Hadronic decay widths are listed in table \ref{tab8} for $J^{PC}$ = $1^{--}$.

\section{Result and discussion}

Some of the exotic states consist of hidden charm quark-antiquark pair with combinations of light quarks or antiquark such as $cq\bar{c}\bar{q}$  where q = u/d and s quarks are being studied. We have analyzed the spectroscopic properties of the tetraquark state using the diquark-diantiquark approach. In this context, we used the relativistic Dirac equation to compute the masses of the confined quarks and antiquarks. Further, the four-body structure of exotic state is factorized into three subsequent two body structure. We have also considered various combinations of the spin and orbital excitations in the calculation of the masses of tetraquark states. In the present calculation, we have first calculated the masses of diquarks and diantiquarks using the same set of parameters that deduced from the meson spectrum. The fitted model parameter are listed in \ref{tab1}. We have considered different potential strength for the inter-cluster interactions of tetraquark systems. The computed masses for meson and diquark/diantiquark states are listed in Table \ref{tab2}. The computed meson masses are in good agreement with experimental results. The spin-spin, spin-orbit and tensor interactions are added to get hyperfine interactions. The S-wave and P-wave masses obtained for the $cq\bar{c}\bar{q}$,  $cs\bar{c}\bar{s}$  and $cs\bar{c}\bar{q}$ are tabulated in Tables \ref{tab3} to  \ref{tab6}. We have also predicted the first radial and orbital excitations of the tetraquark states. Our computed results are compared with available theoretical results. Status of some of the experimental exotic states are identified, and their structural composition is given in Table \ref{tab7}. In view of our present results, we discuss below the status of some of the observed exotic states .\\

\textbf{$\bullet$$X(4500)$ state:}\\

According to our analysis, X(4500) state is a first radial excited state with quark content $cs\bar{c}\bar{s}$. The experimental mass of this state is $4506 \pm 11 ^{+12}_{-15}$ MeV. Our predicted mass of 4505 MeV is close to it. Recently, Zhi-Gang Wang has tentatively assigned X(4500) as the first radial excited state of the axial-vector-diquark-axial-vector-antidiquark $cs\bar{c}\bar{s}$ tetraquark state \cite{4500_zgw}. However, the predicted mass of this state not enough to interpret this state as an exotic state, we need to look for its decay properties also.\\

In the last few years, many charmonium-like $1^{--}$  states have been observed experimentally. The status of these states are still not known and these states are having different structural  properties than expected conventional charmonium states. Here we have tried to forecast the status of some of the $J^{PC}$ = $1^{--}$ states.\\

\textbf{$\bullet$$\psi(4230)$ and $\psi(4260)$ states:}\\

It is clearly observed that the masses of $\psi(4230)$ and $\psi(4260)$ states are very close to each other, but these states are not the same. In 2017, X. Y. Gao, C. P. Shen and C. Z. Yuan et al. have extracted a narrow resonance at Y(4230) and a broad resonance at Y(4260). According to Long-Cheng Gui et al., it is difficult to consider Y(4230) and Y(4260) as pure charmonium states \cite{lcg}. In our previous work, we have discussed S-D wave admixture for Y(4260) state, but the unconfirmed status of this state has motivated us to look for the exotic structure of this state. In our earlier work, We have predicted that Y(4260) is an admixture of $3^3D_1$ and $3^3S_1$ states having decay width 0.258 keV \cite{epjc}. On the other hand, BESIII \cite{bes} also suggested that Y(4260) is not a simple peak. This state is a combination of two resonance Y(4220) and Y(4330) \cite{bes}. In latest PDG-2018 \cite{pdg2018} Y(4230) and Y(4260) are renamed as $\psi(4230)$ and $\psi(4260)$ respectively. According to present analysis, $\psi(4230)$ is a pure $cq\bar{c}\bar{q}$ state having mass 4217  MeV and it$'$s radiative and leptonic decay widths are $\Gamma_{(\psi(4230 \rightarrow J/\psi \gamma)}$ and $\Gamma_{e^+e^-}$     1.932 MeV and 0.391 keV respectively. According to Segovia et al.  the Y(4260) is not pure charmonium state
into the charmonium family in Ref.\cite{js2008}.
In our previous study of bottom tetraquarks, we have calculated admixture of two P-waves of tetraquark systems \cite{sjp2}. The same formalism is used to obtain a mass of $\psi(4260)$ state.

\begin{equation}
M(\psi(4260))=\frac{1}{2}(4217)+\frac{1}{2}(4308) = 4263
\end{equation}

According to present analysis, $\psi(4260)$ state is an admixture of  $^1P_1$ and $^5P_1$  $cq\bar{c}\bar{q}$ states having $J^{PC}$ = $1^{--}$. Predicted radiative and leptonic decay widths the admixed state are 1.959  MeV and 0.374 keV respectively.\\

\textbf{$\bullet$$\psi(4360)$ and $\psi(4390)$ states:}\\

In 2007, the Belle collaboration have observed two states Y (4360) and Y (4660) in the process $e^+e^-$ $\rightarrow$ $\gamma_{ISR} \pi^+ \pi^-\psi'$ \cite{4360-2}. In 2016, the BES collaboration had measured the cross-section of  $e^+e^-$ $\rightarrow$ $\pi^+\pi^- J/\psi$ at the centre of mass energies from 3.77 to 4.60 GeV and observed
two resonances. One resonance has a mass of 4222.0 $\pm$ 3.1 $\pm$ 1.4 MeV which agrees with the mass of Y(4260), and other have a mass of 4320.0 $\pm$ 10.4 $\pm$ 7.0 MeV which again agrees with the mass of Y(4360) \cite{4360-1}.  L. Maiani et al. have used the type-II diquark - antidiquark model and considered the state Y (4360) as the first radial excitations of the state Y (4008) \cite{4360-3}. In their calculations, they have excluded the spin-spin interactions between the quarks and antiquarks. According to Ali et al. the ground state assignment for tetraquark states having L=1 is Y (4220), Y (4330),
Y (4390), Y (4660). Ali et al. have not considered Y (4008), Y (4260),
Y (4360), Y (4660) states that as an exotic states \cite{4360-4}.
 Li and Chao have used the nonrelativistic screened potential model and considered $\psi(4360)$ as the $\psi(3D)$ charmonium state in Ref. \cite{601a}. In Ref. \cite{797a}, Ding, Zhu, and Yan also interpreted the $\psi(4360)$ state as a $3^3D_1$ pure charmonium state they have used the flux tube model to evaluate its leptonic widths, E1 transitions, M1 transitions and the open flavor strong decays. This $\psi(4360)$ state is also interpreted as  a tetraquark state in Ref \cite{512a,796a}, a hadrocharmonium state in Ref \cite{760a}, charmonium hybrid state \cite{737a} and also referred  as a baryonium state in Ref \cite{239a,758a}. $\psi(4360)$ may be the mixture of the pure $D_1\bar{D}^*$
and $D'_1\bar{D}^*$ molecular states. Considering this mixing effect, the decay modes $J/\psi\sigma$, $J/\psi f_0(980)$, $\psi(2S)\sigma$, $\psi(1^3D_1)\sigma$ and $\psi(2S)f_0(980)$ are also suppressed in heavy quark symmetry limit \cite{lm2015}. Thus, the assumption for the molecule structure of the Y(4360) is not possible \cite{lm2015}.
In our previous work we have considered Y(4360) as a mixture
 of $4^3S_1$ and $4^3D_1$ states having  $\Gamma_{e^+e^-}$ = 0.431
  keV but it's experimentally unconfirmed structure leads to investigate
  it's exotic structure. The latest PDG-2018 has renamed Y(4360) as $\psi(4360)$
  state \cite{pdg2018}
According to present analysis, this state is a possible candidate
 of $cs\bar{c}\bar{q}$ tetra quark state. The radiative decay width is 2.009 MeV
 and if we consider this state as an exotic state then its leptonic decay width is 0.369 keV.\\

In Ref. \cite{4360-4}, Ali et al. have considered Y (4390) state as an exotic tetraquark state having L = 1 and very recently Zhi-Gang Wang has predicted Y (4390) as the pure vector tetraquark states. According to present analysis, $\psi(4390)$ is a tetraquark state having quark composition $cs\bar{c}\bar{q}$. The predicted mass of this state is 4404 MeV. We have also predicted its radiative and leptonic decay widths as 2.030 MeV and 0.395 keV. Our predicted status for this state is in accordance with other predictions.\\

\textbf{$\psi(4415)$ state:\\}

The $\psi(4415)$ state is the heaviest and well-established charmonium-like state. This state was first discovered by the Mark I \cite{mark1} in 1976 and DASP \cite{dasp} in 1978 collaborations. The Crystal Ball \cite{tcb} and BESII \cite{bes2} have measured $e^+e^-$ annihilation cross section in the $\psi(4415)$ region. In 2008, G. Pakhlova et al. (Belle collaboration) have observed $\psi(4415)$ $\rightarrow$ $D\bar{D^*_2}$ (2460) decay \cite{gp2008}. In Ref \cite{lph2014} L.P. He et al have suggested that  $\psi(4415)$ cannot be treated as the $\psi(4S)$. Segovia et al have considered  $\psi(4415)$ as the 3D state of $c\bar{c}$ \cite{js2008}. According to present analysis, $\psi(4415)$ is a pure $cs\bar{c}\bar{s}$ state having mass 4416 MeV. \\

\begin{table*}
\begin{center}
\tabcolsep3pt
   \small
\caption{Hadronic decay widths of $1^{--}$ states} \label{tab8}
\begin{tabular}{c| c |c|c|c|c}
\hline\hline
    &    $\psi(4230)$    &    $\psi(4260)$    &    $\psi(4360)$    &    $\psi(4390)$    &    $\psi(4415)$    \\
\hline
Masses (Our) (MeV)    &    4217&    4263    &    4359&    4404&    4416    \\
\hline
Width (MeV)    &        &        &        &        &            \\
$\Gamma$($\psi$ $\rightarrow$ $D^0\bar{D^0})$    &    3.014    &    7.225    &    18.875    &    25.306    &    27.104        \\
$\Gamma$($\psi$ $\rightarrow$ $D^0\bar{D^{0*}})$     &    2.420    &    0.608    &    1.801    &3.914    &    4.557    \\
$\Gamma$($\psi$ $\rightarrow$ $D^{0*}\bar{D^{0*}})$    &    59.297    &    42.028    &    11.736    &2.428    &    0.859        \\
    &        &        &        &        &            \\
$\Gamma$($\psi$ $\rightarrow$ $D_s\bar{D_s})$    &    32.729    &    24.366    &    10.181    &5.214    &    4.098    \\
$\Gamma$($\psi$ $\rightarrow$ $D_s\bar{D_s{*}})$    &    31.133    &    25.648    &    15.680    &    11.707    &    10.726        \\
\hline\hline
\end{tabular}
\end{center}
\end{table*}

\begin{table*}
\begin{center}
\tabcolsep3pt
   \small
\caption{Branching ratio of  $1^{--}$ states} \label{tab9}
\begin{tabular}{c|c|c|c}
\hline\hline
States    &    $\psi(4260)$    &    $\psi(4360)$    &    $\psi(4415)$    \\
\hline\hline
$\textit{BR}$($\psi$ $\rightarrow$ $D^0\bar{D^0})$    &    0.1313    &    0.1966    &    0.4371\\
$\textit{BR}$($\psi$ $\rightarrow$ $D^0\bar{D^{0*}})$     &    0.0110    &    0.0187 &    0.0735\\
$\textit{BR}$($\psi$ $\rightarrow$ $D^{0*}\bar{D^{0*}})$    &    0.7641    &    0.1222    &    0.0138    \\
 &        &        &        \\
$\textit{BR}$($\psi$ $\rightarrow$ $D_s\bar{D_s})$    &    0.4430&    0.1060    &    0.0660    \\
$\textit{BR}$($\psi$ $\rightarrow$ $D_s\bar{D_s{*}})$    &    0.4663    &    0.1633    &    0.1730\\
\hline\hline
\end{tabular}
\end{center}
\end{table*}

In the present study, we have calculated the hadronic decay width
of $J^{PC}$ = $1^{--}$ states using eq. (59), (60) and (61).
The  width of $\psi$ $\rightarrow$ $D^0\bar{D^0}$, $\psi$ $\rightarrow$ $D^0\bar{D^{0*}}$,
$\psi$ $\rightarrow$ $D^{0*}\bar{D^{0*}}$, $\psi$ $\rightarrow$ $D_s\bar{D_s}$  and
$\psi$ $\rightarrow$ $D_s\bar{D_s{*}}$
decays are not measured experimentally yet but using theoretical formalism
we have predicted the width and branching ration of these decays. In the present calculation
of decay widths and branching ratios, we have adopted effective coupling from Ref. \cite{rsa2004}. The calculated hadronic decay
widths and branching ratios are listed in Table \ref{tab8} and \ref{tab9}. States $\psi(4230)$ and $\psi(4390)$ are omitted from meson summary table in PDG
(Particle Data Group) 2018 \cite{pdg2018} but other available theoretical approaches
suggested us to investigate these state. We have calculated hadronic decay widths
of the different channel for these states which are listed in Table $\ref{tab8}$ and as the full width of these states are not available experimentally we have not predicted their branching ratios.\\

We have already discussed the status of very controversial $\psi(4260)$ state as a
diquark-diantiquark state but the mass, radiative decay and leptonic decay are not enough
to resolve the structure of this $1^{--}$ state. According to latest PDG 2018,
$\psi$ $\rightarrow$ $D^0\bar{D^0}$, $\psi$ $\rightarrow$ $D^0\bar{D^{0*}}$,
$\psi$ $\rightarrow$ $D^{0*}\bar{D^{0*}}$, $\psi$ $\rightarrow$ $D_s\bar{D_s}$ and
$\psi$ $\rightarrow$ $D_s\bar{D_s{*}}$  decays are not
seen for $\psi(4260)$ state experimentally but its full width is measured around
$55 \pm 19$ MeV. Our computed hadronic decay widths for $D_s\bar{D_s}$ and $D_s\bar{D_s{*}}$  are higher than it is expected. So, according to present calculation
these are not possible decays for $\psi$ (4260) state. Hadronic decay widths and Branching
Rations for $\psi$ (4260) to $D^0\bar{D^0}$, $D^0\bar{D^{0*}}$ and $D^{0*}\bar{D^{0*}}$ decays are listed in Table
$\ref{tab8}$ and $\ref{tab9}$. Experimentally, we don$'$t have much information about the state $\psi$(4360) but the full
the width and leptonic decay width of this state are measured \cite{pdg2018}.
The hadronic decays of this state are not even seen experimentally but the full width of
96 $\pm$ 7 suggested us to calculate other decays. According to present analysis
the possible decays for $\psi$(4360) are $D^0\bar{D^0}$, $D^0\bar{D^{0*}}$, $D^{0*}\bar{D^{0*}}$,
 $D_s\bar{D_s}$ and
$D_s\bar{D_s{*}}$. The present study has been able to identify many of the observed $1^{--}$
states with their structural compositions. However, still more experimental
support is required to determine the spin parity and quark configuration
of other unknown states. We also look forward to the experimental data on
  $cc\bar{s}\bar{s}$, $cc\bar{q}\bar{q}$, $cq\bar{c}\bar{s}$ states.\\

\section*{Acknowledgement} We acknowledge the financial support from DST-SERB, India (research
Project number: SERB/F/8749/2015-16)

%

\begin{thebibliography}{}
\bibitem{belle} S. K. Choi et al. (Belle Collaboration), Phys. Rev. Lett. \textbf{91}, 262001 (2003).
\bibitem{rlj} R. L. Jaffe, Phys. Rev. D \textbf{15}, 267 (1977).
\bibitem{imb}I. M. Barbour and D. K. Ponting, Nucl. Phys. B  \textbf{149}, 534-546  (1979).
\bibitem{lh}Heller, L.; Tjon, J. A.,  Phys. Rev. \textbf{32}, 3, 755-763 (1985).
\bibitem{bsb} B. Silvestre-Brac, C. Semay, Z. Phys. C \textbf{59}, 457-470 (1993).
\bibitem{bhaduri}R.K. Bhaduri, L.E. Cohler, Y. Nogami, Nuovo Cimento A \textbf{65}, 376, (1981) .
\bibitem{p}M. Ablikim et al. [BESIII Collaboration], Phys. Rev. Lett.\textbf{110}, 252001 (2013).
\bibitem{q} Z. Q. Liu et al. [Belle Collaboration], Phys. Rev. Lett. \textbf{110}, 252002 (2013).
\bibitem{r} T. Xiao, S. Dobbs, A. Tomaradze and K. K. Seth, Phys. Lett. B \textbf{727}, 366 (2013),
\bibitem{s} M. Ablikim et al. [BESIII Collaboration],  Phys. Rev. Lett. \textbf{115}, 112003
(2015).
\bibitem{z} R. Mizuk et al. [Belle Collaboration], Phys. Rev. D \textbf{78}, 072004 (2008).
\bibitem{t} K. Chilikin et al. [Belle Collaboration], Phys. Rev. D \textbf{90}, 11, 112009 (2014).
\bibitem{jw}Jing Wu, Xiang Liu, Yan-Rui Liu, and Shi-Lin Zhu, Phys. Rev. D \textbf{99}, 014037 (2019).
\bibitem{zgw} Zhi-Gang Wang, Eur. Phys. J. C  \textbf{79}: 29 (2019).
\bibitem{jrz}Jian-Rong Zhang, Jing-Lan Zou, Jin-Yun Wu, Chin. Phys. C \textbf{42}, 043101, (2018).
\bibitem{bc} B Chakrabarti et al, Phys. Scr. \textbf{61}, 49, 2000 (1999).
\bibitem{epjc}T. Bhavsar, M. Shah and P. C. Vinodkumar,Eur. Phys. J. C \textbf{78}, 227 (2018).
\bibitem{epj}Manan Shah, Bhavin Patel, P. C. Vinodkumar,Eur. Phys. J. C \textbf{76}, 36 (2016).
\bibitem{prd2014}Manan Shah, Bhavin Patel and P. C. Vinodkumar, Phys. Rev. D \textbf{90}, 014009 (2014).
\bibitem{prd2016}Manan Shah, Bhavin Patel and P. C. Vinodkumar, Phys. Rev. D  \textbf{93}, 094028 (2016).
\bibitem{aruldhas} Quantum mechanics, second edition, G. Aruldhas.
\bibitem{greiner} Relativistic Quantum mechancs wave eqautions by W.Greiner.
\bibitem{mna}Muhammad Naeem Anwar, Jacopo Ferretti, Feng-Kun Guo, Elena Santopinto, Bing-Song Zou, Eur. Phys. J. C \textbf{78}, 647 (2018)
\bibitem{vaghmare} Fundamentals of Quantum mechanics by Y.R.Waghmare
\bibitem{bsrajput} Advance quantum mechanics by B.S. Rajput
\bibitem{amp1} A. P. Monteiro, K. B. Vijaya Kumar,  Natural science \textbf{2},  1292 (2010).
\bibitem{mns}Manan Shah,Bhavin Patel and P C Vinodkumar, Proceedings of the DAE Symp. on Nucl. Phys. \textbf{58}, (2013).
\bibitem{ebert}D.ebert, R.N. Faustov and V. O. Galkin,  phys. Rev. D. \textbf{79}, 114029 (2009).
\bibitem{pcv} P. C. vinodkumar, K. B. Vijayakumar and S. B. Khadkikar, Pramana- J. phys. \textbf{39}, 47 (1992).
\bibitem{ebert2008}D.Ebert, R.N.Faustov and V.O.Galkin, Phys. Lett.B, \textbf{634}, 214 (2006) .
\bibitem{epja2014} Smruti Patel, Manan Shah and P.C. Vinodkumar,Eur. Phys. J. A \textbf{50}, 131  (2014)
\bibitem{hwk}Hong-Wei Ke, Xue-Qian Li and Yan-Liang Shi, Phys. Rev.  D \textbf{87}, 054022 (2013).
\bibitem{jjs}J. J. Sakurai, Currents and Mesons, University of Chicago Press, Chicago, 1969.
\bibitem{lm_2004}L. Maiani, F. Piccini, A. D. Polosa and V. Riquer, Phys. Rev. Lett. \textbf{93}, 212002 (2004)
\bibitem{ar}A. Rehman, arXiv:1109.1095v1
\bibitem{pad}Polosa A. D., Riv. Nuovo Cim., N \textbf{23}, 11 (2000).
\bibitem{js2008}J. Segovia, A.M. Yasser, D.R. Entem, F. Fernandez, Phys. Rev. D \textbf{78}, (2008) 114033
\bibitem{sjp2}S Patel, P C Vinodkumar,  Eur. Phys. J. C  \textbf{76}, 7, 1, (2016)
\bibitem{ahmedali} A. Ali, arXiv:1108.2197v1 [hep-ph]
\bibitem{pdg2018}M.Tanabashi et al. (Particle Data Group), Phys. ReV. D \textbf{98}, 030001 (2018).
\bibitem{zhu2016}Ruilin Zhu, Phys. Rev. D \textbf{94}, 054009 (2016)
\bibitem{rsa2004}R.S.Azevedo and M.Nielsen, Brazilian Journal of Physics, \textbf{34}, 1, (2004).
\bibitem{4500_zgw} Zhi-Gang Wang, Eur. Phys. J. C \textbf{77}, 78 (2017) .
\bibitem{lcg}Long-Cheng Gui, Long-Sheng Lu, Qi-Fang Lu, Xian-Hui Zhong, and Qiang Zhao, arXiv:1801.08791v2 [hep-ph] (2018).
\bibitem{bes}M. Ablikim et al., BESIII Collaboration, Phys Rev. Lett. \textbf{118}, 092001 (2017).
\bibitem{4360-2}X.L. Wang et al., Phys. Rev. Lett. \textbf{99}, 142002 (2007).
\bibitem{4360-1}M. Ablikim et al., Phys. Rev. Lett. \textbf{118},092001 (2017).
\bibitem{4360-3}L. Maiani, F. Piccinini, A.D. Polosa, V. Riquer, Phys. Rev. D \textbf{89}, 114010 (2014).
\bibitem{4360-4}A. Ali, L. Maiani, A.V. Borisov, I. Ahmed, M. Jamil Aslam, A.Y.
Parkhomenko, A.D. Polosa, A. Rehma, Eur. Phys. J. C \textbf{78}, 29 (2018).
\bibitem{601a} B.-Q. Li, K.-T. Chao, Phys. Rev. D \textbf{79}, (2009) 094004.
\bibitem{797a}G.J. Ding, J.J. Zhu, M.L. Yan,  Phys. Rev. D \textbf{77}, 014033  (2008).
\bibitem{512a}  D. Ebert, R.N. Faustov, V.O. Galkin, Eur. Phys. J. C \textbf{58},  399 (2008).
\bibitem{796a}    P. Zhou, C.R. Deng, J.L. Ping,  Chin. Phys. Lett. \textbf{32},  101201 (2015) 
  \bibitem{760a}    X. Li, M.B. Voloshin, Modern Phys. Lett. A \textbf{29} , 1450060 (2014).
   \bibitem{737a}  C.-F. Qiao, L. Tang, G. Hao, X.Q. Li, J. Phys. G \textbf{39}, 015005  (2012)
 \bibitem{239a}  Y.D. Chen, C.F. Qiao, Phys. Rev. D \textbf{85}, 034034, (2012)
\bibitem{758a}  C.F. Qiao,  J. Phys. G \textbf{35},  075008  (2008)
\bibitem{lm2015}Li Ma, Wei-Zhen Deng, Xiao-Lin Chen, Shi-Lin Zhu, arXiv:1512.01938
\bibitem{mark1} J. Siegrist et al. (Mark-1 Collaboration), Phys. Rev. Lett. \textbf{36},  700 (1976).
\bibitem{dasp} R. Brandelik et al. (DASP Collaboration), Phys. Lett. B \textbf{76},361 (1978).
\bibitem{tcb} A. Osterheld et al. (Crystal Ball Collaboration), SLAC Report No. SLAC-PUB-4160, (1986)
\bibitem{bes2} J. Z. Bai et al. (BES Collaboration), Phys. Rev. Lett. \textbf{88}, 101802 (2002).
\bibitem{gp2008}G. Pakhlova et al. (Belle Collaboration),Phys. Rev. Lett. \textbf{100},  062001 (2008)
\bibitem{lph2014}L.P. He, D.Y. Chen, X. Liu, T. Matsuki, , Eur. Phys. J. C \textbf{74 }, 3208 (2014) 
\end{thebibliography}
%

\end{document}